\documentclass[twoside,fleqn]{ActaStyle}
\usepackage[dvips]{graphicx}
\usepackage{times,cite}
\usepackage{mathrsfs}
\usepackage{amsmath}    % need for subequations
\usepackage{graphicx}   % need for figures
\usepackage{verbatim}   % useful for program listings
\usepackage{color}      % use if color is used in text
\usepackage{subfigure}  % use for side-by-side figures
\usepackage{hyperref}   % use for hypertext links, including those to external documents and URLs
\usepackage{rotating}
\usepackage{epic,eepic}

\def\authorheading{}
	\def\shorttitle{Driven Jaynes-Cummings system}
	
\begin{document}
	\pagerange{1}{13}   %(paper has 13 pages)	

	\title{Some features of the Driven Jaynes-Cummings system}

	\author{B. Tuguldur\email{tuguldurb@num.edu.mn} and Ts. Gantsog}
              {School of Physics and Electronics, National University of Mongolia}

	\abstract{We investigate a generalized Jaynes-Cummings model with the external field driving the cavity mode. The numerical results for the dynamics of the atom and the cavity field mode are given. Approximated solutions to the Schr$\ddot{\text{o}}$dinger equation are supplied, which allow us to predict the kinematics of the $Q(\alpha,\alpha^*)$ function in the complex plane.}

\section{Introduction}
The Jaynes-Cummings model (JCM) \cite{jaynes} is one of the simplest models describing the interaction of light with matter, where a single two-level atom interacts with a single mode of quantized radiation field in the electric dipole and rotating wave approximations. This model is of fundamental importance to the field of quantum optics; many interesting features have been predicted for both the atomic variables and the statistical properties of the field through the years, beginning with the well-known phenomena of collapses and revivals of the atomic population inversion oscillation \cite{eberly}. Numerous extensions of the JCM have been considered \cite{alsing,deb,gerry} and many experiments reported \cite{rempe,diedrich,thompson,childs,brune}.

One of the most interesting features of the JCM is that if the atom is initially prepared in its upper state and if the cavity field is initially in the coherent state, then the quantized field evolves into an almost pure state at half of the atomic-revival time \cite{eiselt,phoenix,gea}. This approximately pure state is equal to a superposition (Schr$\ddot{\text{o}}$dinger cat) state composed of two states of light having the same amplitude, but opposite phase. The amplitude of the component states is approximately equal to the amplitude of the initial coherent state of the field mode \cite{buzek}.

Buzek et al. \cite{buzek1} showed that by driving the atom with the external classical field, superposition states of the quantized cavity mode with arbitrary amplitudes and phases of component states can be produced. Recently, Gea-Banacloche and coworkers \cite{gea1} have verified the existence of maximally entangled state in externally driven JCM. Optical cavities with atoms have been proposed for quantum information processing \cite{pellizzari,pellizzari1}.

In this paper we study the dynamics of the JCM when a quantized cavity mode is pumped continuously by an external classical field. In Sec.2 we derive the eigenstates and eigenenergies of the driven Jaynes-Cummings system. Using these results we find the time dependent state vector of the system for a given initial condition. In Sec.3 the dynamics of the atomic inversion, mean photon number and the phase space distribution function $Q(\alpha,\alpha^*)$ are examined. In Sec.4 we give approximate solutions that enable us to explain the features observed. Finally in Sec.5 we summarize our results.
\section{Driven Jaynes-Cummings System}
We consider the driven Jaynes-Cummings system with the external field driving the cavity mode. The interaction Hamiltonian of the system in the interaction picture is given by [3]
\begin{equation}
\hat{H}_I=g(\hat{a}\hat{\sigma}_++\hat{a}^\dagger\hat{\sigma}_-)+\mathscr{E}(\hat{a}^\dagger e^{i\phi}+\hat{a} e^{-i\phi}),\label{hhh}
\end{equation}
where $g$ is the coupling constant between the atom and the cavity mode; $\mathscr{E}$ is the amplitude of the driving field; $\phi$ is the phase of the classical field; $\hat{a}^\dagger$, $\hat{a}$ are creation and annihilation operators for the cavity mode; and $\hat{\sigma}_+$ and $\hat{\sigma}_-$ are atomic pseudospin operators. We assumed that the driving field is in resonance with the atom and the cavity mode.
The steady state solution of the Schr$\ddot{\text{o}}$dinger equation for the Hamiltonian (1) is possible for $\mathscr{E}<g/2$. The quasieigenenergies and the corresponding eigenstates are given by \cite{alsing}
\begin{equation}
\begin{array}{l}
E_0=0,\\
\displaystyle{E_{n1}=g\sqrt{n}\left[1-\left(\frac{2\mathscr{E}}{g}\right)^2\right]^{3/4},\quad n=1,2,3,...,}\\
\displaystyle{E_{n2}=-g\sqrt{n}\left[1-\left(\frac{2\mathscr{E}}{g}\right)^2\right]^{3/4},\quad n=1,2,3,...}
\end{array}
\end{equation}
and
\begin{subequations}\label{steady}
\begin{multline}
\vert\psi_{En1}\rangle=c_p(S(\eta)D(\beta_{n+})\vert n-1\rangle+L_+ S(\eta)D(\beta_{n+})\vert n\rangle)\vert -\rangle\\
+c_p(L_+S(\eta)D(\beta_{n+})\vert n-1\rangle+e^{2i\phi} S(\eta)D(\beta_{n+})\vert n\rangle)\vert +\rangle,
\end{multline}
\begin{multline}
\vert\psi_{En2}\rangle=c_p(S(\eta)D(\beta_{n-})\vert n-1\rangle-L_+ S(\eta)D(\beta_{n-})\vert n\rangle)\vert -\rangle\\
+c_p(L_+S(\eta)D(\beta_{n-})\vert n-1\rangle-e^{2i\phi} S(\eta)D(\beta_{n-})\vert n\rangle)\vert +\rangle,
\end{multline}
\end{subequations}
where if quasienergies are positive (negative) $E_{n1}>0$ ($E_{n2}<0$) then $\displaystyle{\beta_{n+}=-\frac{2\mathscr{E}}{g}\sqrt{n}e^{i\phi}}$ ($\displaystyle{\beta_{n-}=\frac{2\mathscr{E}}{g}\sqrt{n}e^{i\phi}}$), $\eta=re^{i\theta}$, $e^{2r}=\sqrt{1-(2\mathscr{E}/g)^2}$, $\theta=2\phi$,
$c_p=(1/2)\sqrt{1-\sqrt{1-(2\mathscr{E}/g)^2}}$  and $L_{+}=-(ge^{i\phi}/2\mathscr{E})(1+\sqrt{1-(2\mathscr{E}/g)^2})$. For an external field amplitude larger than $\mathscr{E}=g/2$, no normalizable steady states exist. This value is the threshold condition for spontaneous dressed-state polarization.

If the atom is initially prepared in its upper state $\vert +\rangle$ and if the cavity field is prepared initially in the coherent state $\vert \alpha_0\rangle$, then the time evolution of the system is given by
\begin{equation}
\vert\psi(t)\rangle=\vert\psi^+(t)\rangle+\vert\psi^-(t)\rangle,\label{time}
\end{equation}
where
\begin{subequations}\label{timeevolution}
\begin{multline}
\vert\psi^+(t)\rangle=\vert c_p\vert^2
\sum_{n=0}^{\infty}(L_+^*\langle n-1;-\beta_{n+};-\eta\vert\alpha_0\rangle+e^{-2i\phi}\langle n;-\beta_{n+};-\eta\vert\alpha_0\rangle)e^{-iE_nt}\\
\left[(L_+\vert\eta;\beta_{n+};n-1\rangle+e^{2i\phi}\vert\eta;\beta_{n+};n\rangle)\vert +\rangle
+(\vert\eta;\beta_{n+};n-1\rangle+L_+\vert\eta;\beta_{n+};n\rangle)\vert -\rangle\right],\label{df1}
\end{multline}
\begin{multline}
\vert\psi^-(t)\rangle=\vert c_p\vert^2
\sum_{n=0}^{\infty}(L_+^*\langle n-1;-\beta_{n-};-\eta\vert\alpha_0\rangle-e^{-2i\phi}\langle n;-\beta_{n-};-\eta\vert\alpha_0\rangle)e^{iE_nt}\\
\left[(L_+\vert \eta;\beta_{n-};n-1\rangle-e^{2i\phi}\vert \eta;\beta_{n-};n\rangle)\vert +\rangle
+(\vert \eta;\beta_{n-};n-1\rangle-L_+\vert\eta;\beta_{n-};n\rangle)\vert -\rangle\right],\label{df2}
\end{multline}
\end{subequations}
and $E_n=g\sqrt{n}(1-(2\mathscr{E}/g)^2)^{3/4}$ and squeezed and displaced Fock states are denoted as $S(\eta)D(\beta_{n\pm})\vert n\rangle=\vert\eta;\beta_{n\pm};n\rangle$. This result is exact and no approximations are made. We shall apply it to study quantum dynamics of the system in the following section. 
\section{Quantum dynamics of the driven JCM}
\subsection{Atomic inversion and the mean photon number of the cavity mode}
\begin{figure}[!ht]
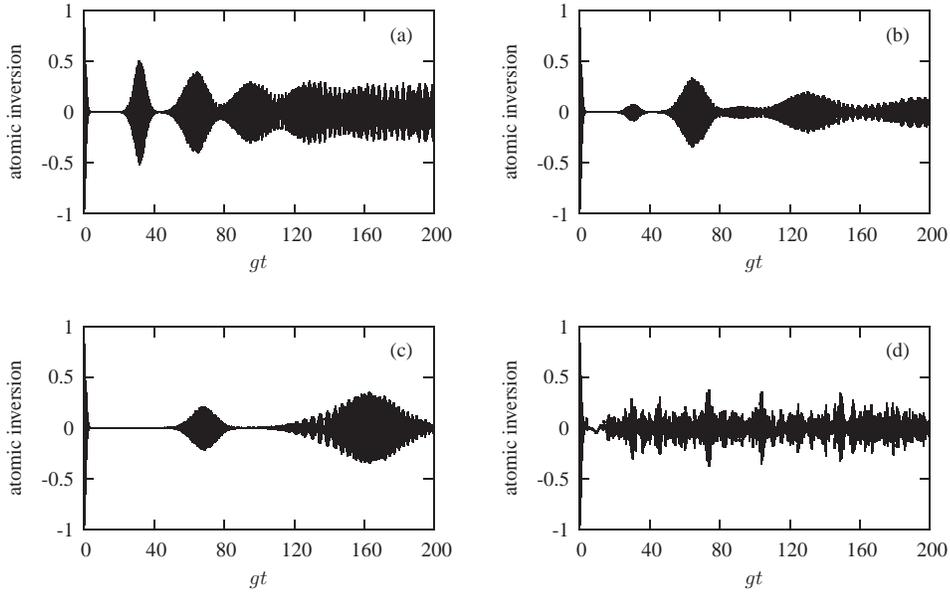

\begin{center}
\subfigure{
\resizebox*{6.5cm}{!}{\input{f1a}}\label{1a}}%
\subfigure{
\resizebox*{6.5cm}{!}{\input{f1b}}\label{1b}}%

\subfigure{
\resizebox*{6.5cm}{!}{\input{f1c}}\label{1c}}%
\subfigure{
\resizebox*{6.5cm}{!}{\input{f1d}}\label{1d}}%
\caption{\label{fig1} The atomic inversion as a function of $gt$ for the atom initially in the upper state and the cavity field in the coherent state $\vert\alpha_0\rangle$. The parameters are $\alpha_0=5$, $g=1$, $\phi=0$ and (a) $\mathscr{E}=0.01$; (b) $\mathscr{E}=0.05$; (c) $\mathscr{E}=0.1$; (d) $\mathscr{E}=2$.}%
\end{center}
\end{figure}

Using explicit expressions for the state vector given by equations \eqref{time} and \eqref{timeevolution} we can study the dynamical properties of the system under consideration. Firstly we calculate the expectation value of the atomic inversion. An illustration of the time evolution of the atomic inversion oscillation in the driven JCM is shown in figure \ref{fig1} for different values of $\mathscr{E}$. 
The revival of the atomic inversion oscillations is a purely quantum-mechanical effect that originates in the discreteness of quantum states of the cavity mode. It is seen from figure \ref{fig1} that with the change of the amplitude $\mathscr{E}$ of the driving field the revival time of the atomic inversion as well as the overall pattern of the time evolution of the atomic inversion oscillation are changed. For larger $\mathscr{E}$ the overall revival and collapse pattern disappear due to the domination of the classical driving field on the quantized cavity field. In the standard JCM ($\mathscr{E}=0$) the total excitation number operator $\hat{R}=\hat{a}^\dagger\hat{a}+1/2(1+\hat{\sigma}_z)$ commutes with the total Hamiltonian and hence represents an integral of motion. As a consequence, the time evolution of the mean photon number of the cavity mode shows the same oscillation as the atomic inversion. However, in the case of driven JCM ($\mathscr{E}\neq 0$) we have $[\hat{R},\hat{H}]\neq 0$, and therefore the total excitation number is not conserved. Physically this corresponds to the fact that the energy is transferred from the classical field to the quantized cavity mode. Using equations \eqref{time} and \eqref{timeevolution} we can numerically calculate the mean photon number of the cavity mode.
\begin{figure}[!ht]
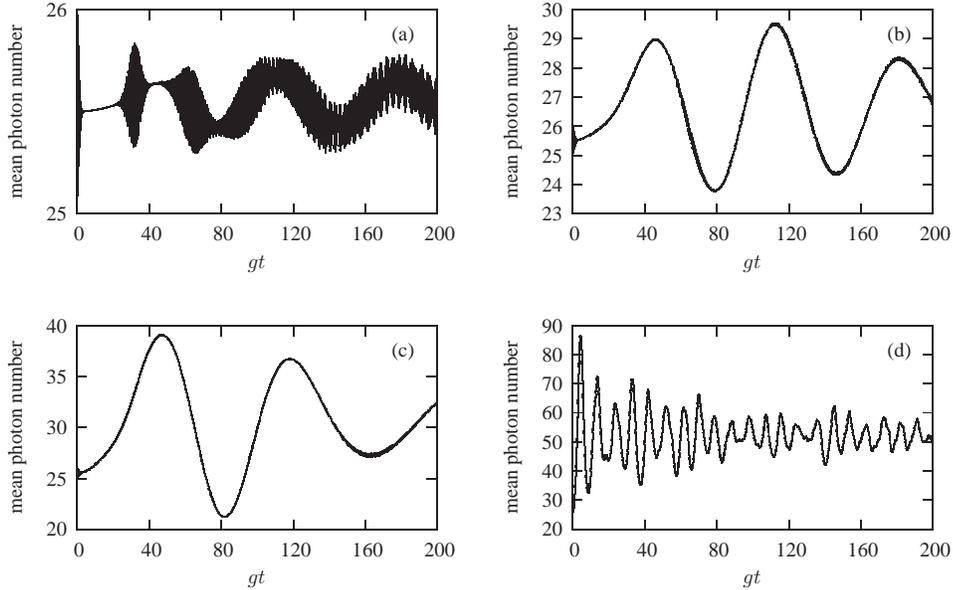

\begin{center}
\subfigure{
\resizebox*{6.5cm}{!}{\input{f2a}}\label{2a}}%
\subfigure{
\resizebox*{6.5cm}{!}{\input{f2b}}\label{2b}}%

\subfigure{
\resizebox*{6.5cm}{!}{\input{f2c}}\label{2c}}%
\subfigure{
\resizebox*{6.5cm}{!}{\input{f2d}}\label{2d}}%
\caption{\label{fig2} The mean photon number as a function of $gt$ for the atom initially in the upper state and the cavity field in the coherent state $\vert\alpha_0\rangle$. The parameters are $\alpha_0=5$, $g=1$, $\phi=0$ and (a) $\mathscr{E}=0.01$; (b) $\mathscr{E}=0.05$; (c) $\mathscr{E}=0.1$; (d) $\mathscr{E}=2$.}%
\end{center}
\end{figure}
In figure \ref{fig2} we plot the mean photon number $\langle\hat{a}^\dagger\hat{a}\rangle$ for $\alpha_0=5$, $\phi=0$ and for different values of $\mathscr{E}$. As we can see from figure \ref{fig2} the time evolution of the mean photon number in the driven JCM exhibits several interesting features. Firstly, for small values of $\mathscr{E}$ (see figure \ref{2a} with $\mathscr{E}=0.01$) a typical collapse-revival pattern corresponding to the time evolution of the inversion can be observed. At the same time we can also observe the pattern of  the global oscillating behavior. With the increase in the classical amplitude $\mathscr{E}$ the time evolution of the mean photon number exhibits stronger global oscillating behavior with small quantum oscillations (revivals) at times corresponding to revivals of the atomic inversion (the amplitude of these oscillations is of the order of unity and therefore they are not clearly seen in our pictures because of the scale used in the y axis).
\begin{figure}
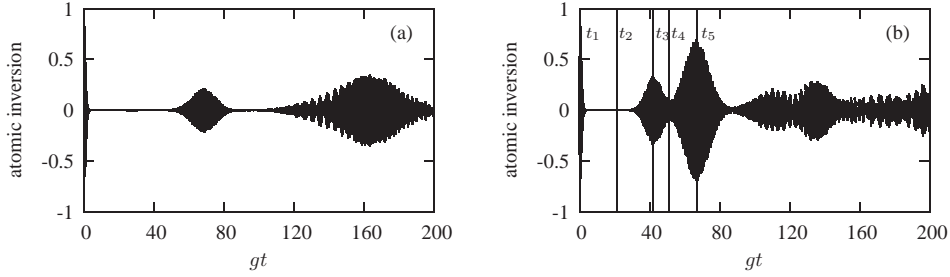

\begin{center}
\subfigure{
\resizebox*{6.5cm}{!}{\input{f3a}}\label{3a}}%
\subfigure{
\resizebox*{6.5cm}{!}{\input{f3b}}\label{3b}}%
\caption{\label{fig3} The atomic inversion as a function of $gt$ for the atom initially in the upper state and the cavity field in the coherent state $\vert\alpha_0\rangle$. The parameters are $\alpha_0=5$, $g=1$, $\mathscr{E}=0.1$ and (a) $\phi=0$ (when $\phi=\pi$, the result is exactly the same as this one); (b) $\phi=\pi/2$ (the vertical bars indicate the times in Fig.5).}%
\label{fig3ab}
\end{center}
\end{figure}
Finishing this part of the paper we turn our attention to the figure \ref{fig3}, where the atomic inversion is plotted for $\alpha_0=5$, $\mathscr{E}=0.1$ and for different values of $\phi$. We can see from the pictures that the revival time of the atomic inversion as well as the overall pattern of its time evolution also depend on the phase of the driving classical field $\phi$.

\subsection{$Q$-function}
Now we turn our attention to the dynamics of the $Q(\alpha,\alpha^*)$ function. We know from the standard JCM ($\mathscr{E}=0$) that, if the initial field is in a coherent state $\vert\alpha_0\rangle$, then during the evolution of the system the single-peaked $Q$-function of the cavity mode bifurcates into two peaks that move in opposite directions around a circular path whose radius equals the amplitude $\alpha_0$ of the initial coherent state \cite{eiselt}. The centre of this circular path is at the origin of the phase space. At one half of the revival time the two peaks have a maximum separation in phase space. Phoenix and Knight \cite{phoenix} and Gea-Banocloche \cite{gea} have shown that at this time the cavity mode (and the atom) returns most closely to the pure state, which is close to a macroscopic superposition composed of two field states that have the same amplitude but opposite phase. The purity of this macroscopic superposition state increases as the amplitude of the cavity mode increases. As the peaks meet at the opposite side of the circle, they produce a revival of the Rabi oscillations.

Buzek et al. \cite{buzek1} have shown that, by driving the atom with the external classical field, one can produce (almost pure) macroscopic superposition states of the cavity field. The amplitudes of the component states can be much larger than the amplitude of the initial cavity field mode. With a proper choice of the phase of the classical driving field, one can produce superposition states with different amplitudes and phases of the component states. 

\begin{figure}[!ht]
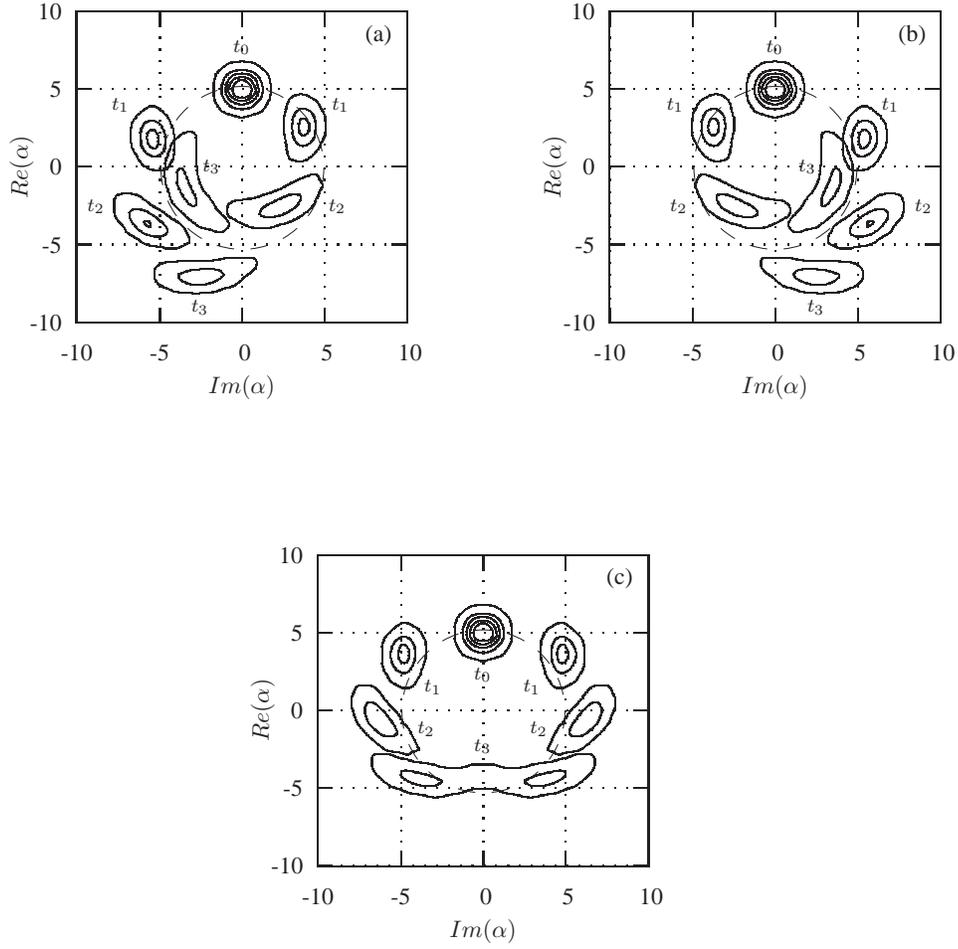

\begin{center}
\subfigure{
\resizebox*{7cm}{!}{\input{f4a}}\label{4a}}%
\subfigure{
\resizebox*{7cm}{!}{\input{f4b}}\label{4b}}%

\subfigure{
\resizebox*{7cm}{!}{\input{f4c}}\label{4c}}%
\caption{\label{fig4} Contour lines of $Q(\alpha,\alpha^*,t)$ in the complex $\alpha$ plane for $\alpha_0=5$, $g=1$, $\mathscr{E}=0.1$ in the initial condition - $\vert\alpha_0\rangle\vert +\rangle$, at the times $gt_0=0$, $gt_1=11.5$, $gt_2=23$, $gt_3=34.5$. (a) $\phi=0$; (b) $\phi=\pi$; (c) $\phi=\pi/2$.}%
\end{center}
\end{figure}

\begin{figure}
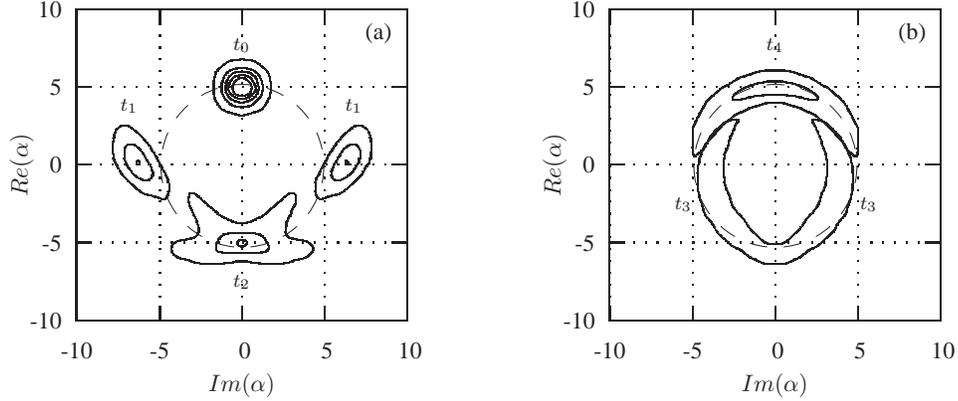

\begin{center}
\subfigure{
\resizebox*{7cm}{!}{\input{f5a}}\label{5a}}%
\subfigure{
\resizebox*{7cm}{!}{\input{f5b}}\label{5b}}%
\caption{\label{fig5} $Q(\alpha,\alpha^*,t)$ function of the cavity mode for the times $gt_0=0$, $gt_1=21$, $gt_2=42$, $gt_3=51$, $gt_4=67$. The parameters are the same as in Fig.3b. The times are indicated by the vertical bars in Fig.3b.}%
\end{center}
\end{figure}

It is interesting to contrast this behavior with what occurs when the cavity quantum mode is driven by an external classical field. In figure \ref{fig4} the contour plots of the $Q(\alpha,\alpha^*)$ function are shown for $\alpha_0=5$, $\mathscr{E}/g=0.1$ and for three different values of the phase $\phi=0$, $\pi$ and $\pi/2$. The most interesting result is that in the course of the time the initial one-peaked function splits into two peaks, which counterrotate on separate circles with different radii depending on the phase of the classical driving field $\phi$. It is clearly seen from the comparison of figure \ref{3b} and figure \ref{fig5} that atomic inversion oscillation shows revival when these two peaks collide. When $\phi=0$ and $\phi=\pi$, the two circles have different radii, and the peaks can collide only at the original site of the circle (see figures \ref{4a} and \ref{4b}). But when $\mathscr{E}/g$ is small enough, the peaks still do have some overlap while moving on different circles (not necessarily on the opposite sides of the circles) and as a result we observe some small revival pattern (see the first small revival in figure \ref{1b}). However, with the increase of $\mathscr{E}/g$, the peaks do not overlap and the revival pattern disappears (see figure \ref{1c}). 

In the next section we will derive an approximate expression for the state vector that enables us to study the kinematics of the $Q(\alpha,\alpha^*)$ function in the complex plane: we will be able to determine radii and centers of two separate circles, angular velocities of rotation of the peaks around these circles, and therefore estimate the revival time of the atomic inversion oscillation.

\section{Approximate solutions}

If $\mathscr{E}$ is small compared to $g$, we can approximate the state vector \eqref{time}-\eqref{timeevolution} of the system to give (see Appendix A) 
\begin{subequations}
\begin{equation}
\vert \psi^+(t)\rangle\approx A^+(t)\vert (\alpha_0+\bar{\beta}_n)e^{-i\omega^+gt}-\bar{\beta}_n\rangle(\vert +\rangle+\vert -\rangle),\label{tr1}
\end{equation}
\begin{equation}
\vert \psi^-(t)\rangle\approx A^-(t)\vert (\alpha_0-\bar{\beta}_n)e^{i\omega^-gt}+\bar{\beta}_n\rangle(\vert +\rangle-\vert -\rangle),\label{tr2}
\end{equation}
\end{subequations}
where $\bar{\beta}_n=(2\mathscr{E}/g)\sqrt{\bar{n}}e^{i\phi}$, and
\begin{subequations}
\begin{multline}
A^+(t)=\vert c_p\vert^2 \vert L_+\vert^2\exp\left(\frac{1}{2}\vert\alpha_0+\bar{\beta}_n\vert^2\right)\\
\exp\left\{\frac{1}{2}[-\bar{\beta}_n(\alpha_0+\bar{\beta}_n)^*e^{i\omega^+gt}+\bar{\beta}_n^*(\alpha_0+\bar{\beta}_n)e^{-i\omega^+gt}]\right\}
\exp\left[\frac{1}{2}(\alpha_0^*\bar{\beta}_n-\alpha_0\bar{\beta}_n^*)\right],\label{an}
\end{multline}
\begin{multline}
A^-(t)=\vert c_p\vert^2 \vert L_+\vert^2\exp\left(\frac{1}{2}\vert\alpha_0-\bar{\beta}_n\vert^2\right)\\
\exp\left\{\frac{1}{2}(\bar{\beta}_n(\alpha_0-\bar{\beta}_n)^*e^{-i\omega^-gt}-\bar{\beta}_n^*(\alpha_0-\bar{\beta}_n)e^{i\omega^-gt})\right\}
\exp\left[\frac{1}{2}(-\alpha_0^*\bar{\beta}_n+\alpha_0\bar{\beta}_n^*)\right],\label{ah}
\end{multline}
\end{subequations}
and (see Appendix B)
\begin{subequations}
\begin{equation}
\omega^+=\frac{[1-(2\mathscr{E}/g)^2]^{3/4}}{2\vert\alpha_0+\bar{\beta}_n\vert},
\end{equation}
\begin{equation}
\omega^-=\frac{[1-(2\mathscr{E}/g)^2]^{3/4}}{2\vert\alpha_0-\bar{\beta}_n\vert}.
\end{equation}
\end{subequations}
It is clear from the above results that $\psi^+$ corresponds to the coherent state rotating along the circle of radius $\vert \alpha_0+\bar{\beta}_n\vert$ centered at $-\bar{\beta}_n$, and  $\psi^-$  corresponds to the coherent state rotating along the circle of radius $\vert \alpha_0-\bar{\beta}_n\vert$ centered at $\bar{\beta}_n$. These states counterrotate in the complex plane. This approximation is valid only if time is less than revival time.

\begin{figure}
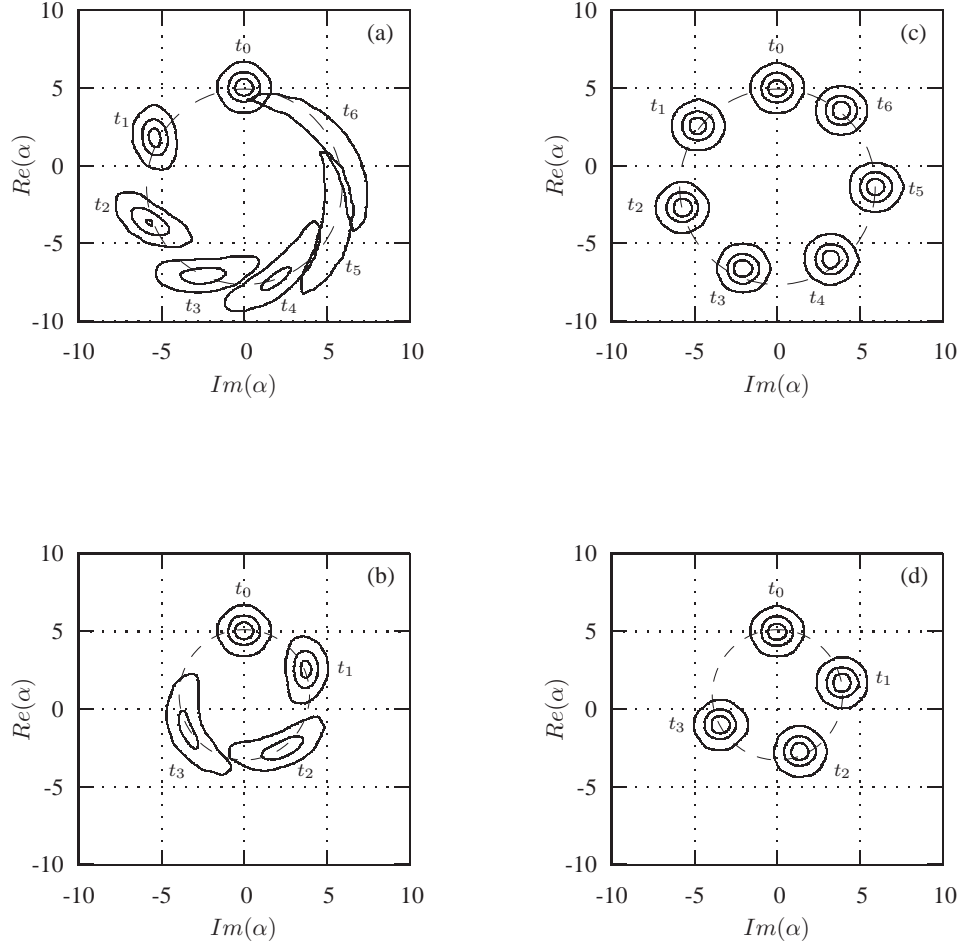

\begin{center}
\subfigure{
\resizebox*{7cm}{!}{\input{f6a}}\label{6a}}%
\subfigure{
\resizebox*{7cm}{!}{\input{f6c}}\label{6c}}%

\subfigure{
\resizebox*{7cm}{!}{\input{f6b}}\label{6b}}%
\subfigure{
\resizebox*{7cm}{!}{\input{f6d}}\label{6d}}%
\caption{\label{fig6} Contour lines of $Q(\alpha,\alpha^*,t)$ in the complex $\alpha$ plane for the initial condition $\vert +\rangle\vert\alpha_0\rangle$ and for $\alpha_0=5$, $g=1$, $\mathscr{E}=0.1$. (a) $\phi=0$, $\psi^+(t)$; (b) $\phi=0$, $\psi^-(t)$; (c) and (d) are the approximated results and parameters are same as (a) and (b) respectively.}%
\end{center}
\end{figure}

The modules of the $A^+$ and $A^-$ are not time dependent. Therefore density operator of the cavity mode takes the form
\begin{equation}
\rho_f=Tr_{atom}(\rho)=2\vert A^+\vert^2\vert\gamma^+(t)\rangle\langle\gamma^+(t)\vert+2\vert A^-\vert^2\vert\gamma^-(t)\rangle\langle\gamma^-(t)\vert,
\end{equation}
where $\rho$ is the density operator for the total system, $\gamma^+(t)=(\alpha_0+\bar{\beta}_n)e^{-i\omega^+gt}-\bar{\beta}_n$, and $\gamma^-(t)=(\alpha_0-\bar{\beta}_n)e^{i\omega^-gt}+\bar{\beta}_n$. Then $Q(\alpha,\alpha^*)$ function can be written as a sum of two counterrotating peaks in the phase space i.e.,
\begin{equation}
Q(\alpha,\alpha^*,t)\sim \vert\langle\alpha\vert\gamma^+(t)\rangle\vert^2+\vert\langle\alpha\vert\gamma^-(t)\rangle\vert^2\label{qq}.
\end{equation}
In figure \ref{fig6} we plotted the time evolution of $Q(\alpha,\alpha^*,t)$ using exact, as well as approximated solutions for $\phi=0$ (figures \ref{6a} and \ref{6b}) and for $\phi=\pi/2$ (figures \ref{6c} and \ref{6d}). The initial condition of the system is $\vert +\rangle\vert\alpha_0\rangle$. The left hand side pictures correspond to the exact solutions and the right hand side pictures correspond to the approximated results. The initial $Q(\alpha,\alpha^*)$ distribution is a Gaussian centered at $\alpha_0$. As we know, in the course of time the initial one-peaked function splits into two peaks. We plotted them in separate Figures. One of them ($\psi^+$) rotates counterclockwise (see figure \ref{6a}) on the circle of radius $\vert\alpha_0+1\vert=6$ ($\bar{\beta}_n\approx 1$) with angular speed $\omega^+=0.0808$, and the other one ($\psi^-$) rotates clockwise (see figure \ref{6b}) on the circle with smaller radius $\vert\alpha_0-1\vert=4$ at angular speed $\omega^-=0.1212$. In this case the centers of circles are shifted respectively down and up in real axis.
\begin{figure}
\begin{center}
\subfigure{
\resizebox*{7cm}{!}{\input{f7a}}\label{7a}}%
\subfigure{
\resizebox*{7cm}{!}{\input{f7c}}\label{7c}}%

\subfigure{
\resizebox*{7cm}{!}{\input{f7b}}\label{7b}}%
\subfigure{
\resizebox*{7cm}{!}{\input{f7d}}\label{7d}}%
\caption{\label{fig7} Contour lines of $Q(\alpha,\alpha^*,t)$ in the complex $\alpha$ plane for the initial condition $\vert +\rangle\vert\alpha_0\rangle$ and for $\alpha_0=5$, $g=1$, $\mathscr{E}=0.1$. (a) $\phi=0$, $\psi^+(t)$; (b) $\phi=0$, $\psi^-(t)$; (c) and (d) are the approximated results and parameters are same as (a) and (b) respectively.}%
\end{center}
\end{figure}
For $\phi=\pi/2$, $\psi^+$ and $\psi^-$ counterrotate on the different circles with the same radius $\vert\alpha_0\pm i\vert=5.099$ at angular speed $\omega^{\pm}=0.0951$. The centers of the circles are shifted horizontally. We can see from figure \ref{fig6} that our approximated solutions give almost precise locations of the peaks. Therefore we can predict the kinematics of the moving peaks and find the times when those peaks collide, which enable us to calculate the revival times of the atomic inversion.

\section{Conclusion}

We have calculated steady states of the generalized Jaynes-Cummings system with the external field driving the cavity mode. The steady states are the superposition of atomic states multiplied by squeezed and displaced Fock states. The calculated steady states are valid only if $2\mathscr{E}/g<1$. In other words, $2\mathscr{E}/g=1$ is a threshold value of the driving field strength and above this value normalizable steady states do not exist. But we numerically calculated the time evolution of normalized states for the arbitrary value of $2\mathscr{E}/g$. Using these results we studied the dynamics of the atomic variables and the statistical properties of the field.
 
Furthermore, we derived approximate solutions for the time evolution of quantum states, which allow us to understand the kinematics of $Q(\alpha,\alpha^*)$ function in the phase space. We found that as time goes on the initial single peaked $Q(\alpha,\alpha^*)$ function splits into two peaks which counterrotate on separate circles of different radii $\vert\alpha_0\pm\bar{\beta}_n\vert$ with different angular velocities $\omega^{\pm}$. Whenever these peaks collide, or even partly overlap, atomic inversion oscillation shows revival.

\appendix
\section{Approximation method}

If we are interested only in the kinematics of the $Q(\alpha,\alpha^*)$ function in the complex plane, but not the shapes of the peaks, then we can make some very simple approximations for the state vectors \eqref{timeevolution}. We restrict our calculations with only first order terms of $\mathscr{E}/g$ ignoring all higher order terms to yield $r\approx 0$, and therefore $\eta\approx 0$, and $S(\eta)\approx 1$ (squeezing operator does not affect the location of the peaks in phase space, it changes only the shape of the peaks). 

Doing same approximation we get  $L_+\approx-(g/\mathscr{E})e^{i\phi}$. Since $\vert L_+\vert$ is much greater than 1, the terms not containing $L_+$ can be neglected in equations \eqref{timeevolution}. Then we get 
\begin{equation}
\vert\psi^+(t)\rangle\approx\vert c_p\vert^2\vert L_+\vert^2\sum_{n=0}^{\infty}(\langle n-1;\beta_{n}\vert\alpha_0\rangle)e^{-iE_nt}
\left(\vert -\beta_{n};n-1\rangle\vert +\rangle+\vert -\beta_{n};n\rangle\vert -\rangle\right),
\end{equation}
\begin{equation}
\vert\psi^-(t)\rangle\approx\vert c_p\vert^2\vert L_+\vert^2\sum_{n=0}^{\infty}(\langle n-1;-\beta_{n}\vert\alpha_0\rangle)e^{iE_nt}
\left(\vert \beta_{n};n-1\rangle\vert +\rangle-\vert\beta_{n};n\rangle\vert -\rangle\right),
\end{equation}
where $\beta_n=(2\mathscr{E}/g)\sqrt{n}e^{i\phi}$. The difficulty to manage above equations is that $\beta_n$ depends on the photon number $n$. The multiplier $2\mathscr{E}/g$ in equation $\beta_n$ is much less than $1$. That leads to $\beta_n\neq 0$ in small interval of $n$. If we change $\beta_n$ with 
\begin{equation}
\bar{\beta}_n=(2\mathscr{E}/g)\sqrt{\bar{n}}e^{i\phi},
\end{equation}
we can easily evaluate state vector
\begin{equation}
\vert\psi^+(t)\rangle\approx\vert c_p\vert^2\vert L_+\vert^2\sum_{n=0}^{\infty}\frac{(\alpha_0+\bar{\beta}_n)^n}{\sqrt{n!}}e^{-iE_nt}
D(-\bar{\beta}_n)\left(\vert n-1\rangle\vert +\rangle+\vert n\rangle\vert -\rangle\right),\label{app1}
\end{equation}
\begin{equation}
\vert\psi^-(t)\rangle\approx\vert c_p\vert^2\vert L_+\vert^2\sum_{n=0}^{\infty}\frac{(\alpha_0-\bar{\beta}_n)^n}{\sqrt{n!}}e^{iE_nt}
D(\bar{\beta}_n)\left(\vert n-1\rangle\vert +\rangle-\vert n\rangle\vert -\rangle\right).\label{app2}
\end{equation}
If $\alpha_0$  is large enough, then $\alpha_0+\bar{\beta}_n$ and $\alpha_0-\bar{\beta}_n$ are also large and there is not much difference between $\vert n-1\rangle$ and $\vert n\rangle$. We can substitute $e^{\mp i\omega^{\pm} nt}$ instead of $e^{\mp iE_nt}$. Therefore, taking into account the relation
\begin{equation}
D(\alpha)D(\beta)=\exp\left[\frac{1}{2}(\alpha\beta^*-\alpha^*\beta)\right]D(\alpha+\beta),
\end{equation}
we get the approximate expressions for the state vectors as
\begin{equation}
\vert\psi^+(t)\rangle\approx A^+(t)\sum_{n=0}^{\infty}\frac{[(\alpha_0+\bar{\beta}_n)e^{-i\omega^+gt}-\bar{\beta}_n]^n}{\sqrt{n!}}\vert n\rangle\left(\vert +\rangle +\vert -\rangle\right),\label{app3}
\end{equation}
\begin{equation}
\vert\psi^-(t)\rangle\approx A^-(t)\sum_{n=0}^{\infty}\frac{[(\alpha_0-\bar{\beta}_n)e^{i\omega^-gt}+\bar{\beta}_n]^n}{\sqrt{n!}}\vert n\rangle\left(\vert +\rangle -\vert -\rangle\right).\label{app4}
\end{equation}

\section{Estimations of $\omega^+$ and $\omega^-$}

Here we make a simple estimation for the rotational angular velocities $\omega^+$ and $\omega^-$ of two separate peaks of the $Q(\alpha,\alpha^*)$ function in the complex plane. First we estimate $\omega^+$. Using equation \eqref{app1}, the reduced density operator for the quantized cavity field can be written as
\begin{equation}
\rho_f^+=\sum^{\infty}_{n,m=0}\frac{(\alpha_0+\bar{\beta}_n)^n}{\sqrt{n!}}\frac{((\alpha_0+\bar{\beta}_n)^*)^m}{\sqrt{m!}}e^{-ikgt(\sqrt{n}-\sqrt{m})}D(-\bar{\beta}_n)\vert n\rangle\langle m\vert D(\bar{\beta}_n),\label{rho}
\end{equation}
where $k=(1-(2\mathscr{E}/g)^2)^{3/4}$. Using a Taylor expansion of $\sqrt{n}$ around $\sqrt{\bar{n}}$  (where $\bar{n}=\vert\alpha_0+\bar{\beta}_n\vert^2$)
\begin{equation}
\sqrt{n}\approx \sqrt{\bar{n}}+\frac{1}{2\sqrt{\bar{n}}}(n-\bar{n})-\dots,
\end{equation}
and ignoring all higher order terms we get
\begin{equation}
\sqrt{n}-\sqrt{m}\approx (n-m)/2\sqrt{\bar{n}}.\label{fff}
\end{equation}                                                
Substituting it into the equation \eqref{rho} for $\rho_f^+$ one can find $\omega^+$ as follows
\begin{equation}
\omega^+=\frac{[1-(2\mathscr{E}/g)^2]^{3/4}}{2\vert\alpha_0+\bar{\beta}_n\vert}.
\end{equation}
Similarly, for $\omega^-$ we get
\begin{equation}
\omega^-=\frac{[1-(2\mathscr{E}/g)^2]^{3/4}}{2\vert\alpha_0-\bar{\beta}_n\vert}.
\end{equation}

\end{document}